\documentstyle[aps%,twocolumn%
,epsfig%
,preprint%
]{revtex}
\begin{document}
\draft
\preprint{Guchi-TP-016}
\date{\today
}
\title{
Non-topological soliton in three-dimensional supergravity
}
\author{Yoshinori~Cho${}^{1}$%
\thanks{e-mail: {\tt
b2669@sty.cc.yamaguchi-u.ac.jp}
% \hfill\break
},
Kenji~Sakamoto${}^{1}$%
\thanks{e-mail: {\tt
b1795@sty.cc.yamaguchi-u.ac.jp}
% \hfill\break
} and
Kiyoshi~Shiraishi${}^{1,2}$%
\thanks{e-mail: {\tt
shiraish@sci.yamaguchi-u.ac.jp}
% \hfill\break
}}
\address{${}^1$Faculty of Science, Yamaguchi University\\
Yoshida, Yamaguchi-shi, Yamaguchi 753-8512, Japan
}
\address{${}^2$Graduate School of Science and Engineering,
Yamaguchi University\\
Yoshida, Yamaguchi-shi, Yamaguchi 753-8512, Japan
}
\maketitle
\begin{abstract}
%%%%%%%%%%%%%%%%%%%%%%%%%%%%%%%%%%%%%%%%%%%%%%%%%%%%%%%%%%%%%%%%%%%%%%
We construct numerical solutions for non-topological solitons in 
three-dimensional $U(1)$-gauged ${\cal N}=2$ supergravity. 
We find the region of the solutions showing 
with the BTZ mass, the angular momentum and the magnetic flux and 
discuss the relation 
among the physical parameters for various values of a cosmological
 constant.
%%%%%%%%%%%%%%%%%%%%%%%%%%%%%%%%%%%%%%%%%%%%%%%%%%%%%%%%%%%%%%%%%%%%%%
\end{abstract}

%\vspace{7mm}
%\vfill
%\eject

%%%%%%%%%%%%%%%%%%%%%%%%%%%%%%%%%%%%%%%%%%%%%%%%%%%%%%%%%%%%%%%%%%%%%%
\section{introduction}
%%%%%%%%%%%%%%%%%%%%%%%%%%%%%%%%%%%%%%%%%%%%%%%%%%%%%%%%%%%%%%%%%%%%%%
%%%%%%%%%%%%%%%%%%%%%%%%%%%%%%%%%%%%%%%%%%%%%%%%%%%%%%%%%%%%%%%%%%%%%%
As is well known, 
three-dimensional pure Einstein gravity is a topological theory 
and has no propagating degrees of freedom. 
The solutions are locally flat apart from conical singularities 
at the location of matter sources in no cosmological constant 
case~\cite{ref1,ref2}. 
Moreover, cosmological gravity 
and supergravity are solvable 
in many cases~\cite{ref3,ref4,ref5,ref6,ref7,ref8,ref9}.

Three-dimensional supergravity has been investigated 
in various situations intensively. 
As particular examples, the non-linear sigma models coupled to 
${\cal N}$-extended supergravity were studied~\cite{ref10}. 
The geometry of the target manifolds parametrized by scalar fields 
were classified completely: 
the target space is Riemanian, K\"{a}hler, and quaternionic 
for ${\cal N}=1,2$ and ${\cal N}=3$ respectively. 
For ${\cal N}=4$ it generally decomposes 
into two separate quaternionic spaces, associated with inequivalent 
supermultiplets. 
For ${\cal N}=5,6,8$ there is a symmetric space for any given 
number of supermultiplets. 
For ${\cal N}=9,10,12$ and $16$ there are only theories based on 
a single supermultiplet associated with coset spaces with the 
exceptional isometry groups $F_4,E_6,E_7$ and $E_8$ respectively. 

Moreover, the three-dimensional maximal (${\cal N}=16)$ gauged 
supergravity were constructed~\cite{ref11,ref12}. 
The duality relation between the gauge fields and the scalar 
field plays an important role in the derivation of the 
ungauged ${\cal N}=16$ supergravity: in order to expose its global 
$E_{8}$ isometry, all vector fields obtained by dimensional 
reduction of the $D=11$ supergravity 
in an eight-torus must be dualized into the scalar fields. 
Moreover, in order for the supergravity to gauge, 
the Chern-Simons term (rather than the Yang-Mills term) 
must be required in the Lagrangian.

Recently the supersymmetric vortex solutions in $U(1)$-gauged 
$D=3,{\cal N}=2$ supergravity whose scalar sector 
is an arbitrary K\"{a}hler manifold with $U(1)$ isometry were 
constructed~\cite{ref16}. 
It has been found that the Einstein equations and the matter 
field equations of the model can be recast 
into a set of self-duality equations for a specific eighth-order 
choice of the Higgs potential which 
reduces to the sixth-order potential of the flat space model 
when the Newton gravitational coupling 
constant is set to zero. 
The Abelian Chern-Simons Higgs model in three-dimensional Minkowski 
space-times and its vortex solutions were studied 
by~\cite{ref13,ref14,ref15}. 
It was shown that the model with a specific sixth-order Higgs 
potential admits topologically stable vortex solutions and 
non-topological soliton solutions with nonzero flux and charge, 
which satisfy the Bogomol'nyi-type or the first order 
self-duality equations~\cite{ref19}. 
In the Bogomol'nyi limit, there is a (first-order) phase 
transition point between the symmetric (non-topological solutions) 
and the asymmetric phase (topological vortex solutions).

In this paper, we construct the non-topological solitons, 
which do not preserve the supersymmetry,
in three-dimensional $U(1)$-gauged ${\cal N}=2$ supergravity 
and find the region where the solutions 
can exist for various value of a cosmological constant. 
A further motivation for studying the non-topological solitons 
is that this model provided us with the simplest model of 
relativistic stars 
as self-gravitating systems. 
Rotating boson stars with large self-coupling constant 
in three dimensions were studied in~\cite{ref17}, for example. 
The study of the non-topological 
soliton will lead to a new aspect of the self-gravitating systems.

The organization of this paper is as follows; 
in the next section we introduce the model 
and the basic ingredients. For the construction of the 
$U(1)$-gauged $D=3$, ${\cal N}=2$ 
supergarvity Lagrangian, we refer the reader to the existing 
literature~\cite{ref10,ref11,ref12,ref16}. 
In section~\ref{sec:3}, we restrict ourselves to the simple 
case of the K\"{a}hler manifold 
and a circularly symmetric ansatz for the metric which 
approaches asymptotically the BTZ 
configuration~\cite{ref18}, which is asymptotically 
anti-de Sitter space (rather than asymptotically flat) 
and does not preserve the supersymmetry in general. 
In this setting we obtain numerically the 
non-topological solitons with zero-winding number case for 
various values of a cosmological constant. 
In the section~\ref{sec:4}, we show the numerical results 
and discuss the relation among 
the physical parameters for various values of a cosmological constant.

%%%%%%%%%%%%%%%%%%%%%%%%%%%%%%%%%%%%%%%%%%%%%%%%%%%%%%%%%%%%%%%%%%%%%%
\section{$U(1)$-gauged $D=3$, ${\cal N}=2$ supergarvity}
%%%%%%%%%%%%%%%%%%%%%%%%%%%%%%%%%%%%%%%%%%%%%%%%%%%%%%%%%%%%%%%%%%%%%%
%%%%%%%%%%%%%%%%%%%%%%%%%%%%%%%%%%%%%%%%%%%%%%%%%%%%%%%%%%%%%%%%%%%%%%
In this section we consider ungauged ${\cal N}=2$ three-dimensional 
supergarvity. Assuming a $U(1)$ 
isometry of the K\"{a}hler potential, we apply the Noether procedure 
to obtain the $U(1)$-gauged 
${\cal N}=2$ supergarvity~\cite{ref10,ref16}. The field content 
of the ungauged theory is the following.

\begin{itemize}
\item
The ${\cal N}=2$ supergravity multiplet
\begin{equation}
 \{e^{a}_{\mu},\psi_{\mu} \}
\end{equation}
contains a graviton~$e^a_{\mu}$~and two gravitini which are 
assembled into one complex spinor $\psi_{\mu}$.

\item
The ${\cal N}=2$ scalar multiplet
\begin{equation}
 \{\phi^{\alpha},\lambda^{\alpha}\}
\end{equation}
contains the fermions 
$(\lambda^{\alpha},\bar{\lambda}^{\bar{\alpha}})$ and scalars 
$(\phi^{\alpha},\bar{\phi}^{\bar{\alpha}})$ with $\alpha=1,\dots,p$, 
respectively in the complex notation. 
The matter sector is obtained by $p$ copies 
of the ${\cal N}=2$ scalar multiplet. 
The scalar fields 
define a K\"{a}hler manifold of real dimension $2p$, 
characterized by its K\"{a}hler potential 
$K(\phi^{\alpha},\bar{\phi}^{\bar{\alpha}})$.
\end{itemize}

It is assumed that the ungauged ${\cal N}=2$ 
supergarvity is invariant under the global $U(1)$ 
isometry. As in the maximally gauged supergravity 
theories~\cite{ref11,ref12}, the Chern-Simons
 term is required, in order to gauge the $U(1)$ 
K\"{a}hler isometry. 
Note that this term is 
topological and hence does not introduce new 
propagating degrees of freedom in the gauged theory.

Moreover, the $g$-dependent terms, 
due to the Chern-Simons terms, give rise to extra terms 
in supersymmetry transformations. 
In order to compensate these terms, the extra Yukawa-type 
bilinear fermionic terms and a scalar potential 
must be added to the Lagrangian.

Then the bosonic parts of the $U(1)$-gauged $D=3$, 
${\cal N}=2$ supergarvity Lagrangian is given by:
\begin{equation}
 {\cal L}=
 \frac{1}{4}eR
 -eG_{\alpha\bar{\alpha}}(\phi,\bar{\phi})
 {\cal D}_{\mu}\phi{\cal D}^{\mu}\bar{\phi}
 -\frac{1}{8}g\epsilon^{\mu\nu\rho}A_{\mu}F_{\nu\rho}-eg^2V,
 \label{Lagrangian}
\end{equation}
where ${D}_{\mu}\phi\equiv(\partial_{\mu}+igA_{\mu})\phi$ 
is a covariant 
derivative and $G_{\alpha\bar{\alpha}}(\phi,\bar{\phi})$ 
denotes the K\"{a}hler 
metric 
$G_{\alpha\bar{\alpha}}(\phi,\bar{\phi})
=\partial_{\alpha}\partial_{\bar{\alpha}}K(\phi,\bar{\phi})$.
 $F_{\mu\nu}$ is an Abelian field strength and $g$ 
is a gauge coupling constant. $V=V(\phi,\bar{\phi})$ 
is a real scalar potential, which will 
be mentioned in the next section.

%%%%%%%%%%%%%%%%%%%%%%%%%%%%%%%%%%%%%%%%%%%%%%%%%%%%%%%%%%%%%%%%%%%%%%
\section{Non-topological soliton}
\label{sec:3}
%%%%%%%%%%%%%%%%%%%%%%%%%%%%%%%%%%%%%%%%%%%%%%%%%%%%%%%%%%%%%%%%%%%%%%
%%%%%%%%%%%%%%%%%%%%%%%%%%%%%%%%%%%%%%%%%%%%%%%%%%%%%%%%%%%%%%%%%%%%%%
We restrict ourselves to the simple case of 
the Lagrangian~(\ref{Lagrangian}) 
considered in the preceding section. 
We adopt a single complex scalar field case, 
that is $p=1$. Moreover 
we assume that the $K$ is a function of $R=|\phi|$ only 
and the K\"{a}hler potential 
is $K(R)=R^2$ in order to obtain the canonical kinetic 
term for the scalar field. 
We thus find that the $G\equiv G_{1\bar{1}}$ is unity.

Then, the eighth-order scalar potential is reduced to:
\begin{equation}
 V=-2g^2[\phi^8-2(2c+1)\phi^6+2(3c^2+2c+b)\phi^4
 -2c(2c^2+c+2b)\phi^2+c^4+b^2+2bc^2],
\end{equation}
where $b$ and $c$ are arbitrary real constant parameters.

We assume the three-dimensional 
metric for a circularly symmetric spacetimes as
\begin{equation}
 ds^2=-e^{-2\delta(r)}\Delta(r)dt^2+\Delta^{-1}(r)dr^2
      +r^2(d\theta^2-\Omega(r)dt)^2,
\end{equation}
where $\delta$, $\Delta$, 
and $\Omega$ are functions of the radial coordinate $r$ only.

For the single complex scalar field, 
we also assume the following dependence 
on the coordinates:
\begin{equation}
 \phi=R(r)e^{-in\theta},
\end{equation}
where $R(r)$ is a function of the radial coordinate $r$ only and $n$ is a 
constant. We impose the boundary condition that $\phi(r)$ approaches zero 
when the radial coordinate $r$ is taken to infinity, 
in order to obtain the non-topological soliton.

For the vector field $A_{\mu}$, we choose the gauge, in order to require 
finite energy solution for the scalar field at radial infinity, in which
\begin{equation}
 A_r=0, \>\>\> A_{\theta}=P(r)+\frac{n}{g},\>\>\> A_t=W(r),
\end{equation}
where $P(r)$ and $W(r)$ are functions of the radial coordinate $r$ only.

Varying the Lagrangian with respect to the gauge field, the scalar field
 and the metric yields equations of motion. 
The equation of motion for the 
Chern-Simons term induces the first order equations
\begin{equation}
 \epsilon^{\mu\nu\rho}F_{\nu\rho}
 =8ire^{-\delta}(\bar{\phi}{\cal D}^{\mu}\phi
 -\phi{\cal D}^{\mu}\bar{\phi}),
\end{equation}
and two of these equations take the form
\begin{eqnarray}
 \partial_rW
 &=&
 -4gr\Delta^{-1}e^{\delta}R^2\Omega(W+\Omega P)
 +4gR^2Pr^{-1}e^{-\delta},\\
 \partial_rP
 &=&
 4gr\Delta^{-1}R^2e^{\delta}(W+\Omega P),
\end{eqnarray}
and the third one ($\mu=r$) is automatically satisfied.

The scalar field equation is reduced to the form:
\begin{equation}
 R''+\left(\frac{\Delta'}{\Delta}+\frac{1}{r}-\delta'\right)R'
 +\frac{g^2e^{2\delta}R}{\Delta^2}
 (W+\Omega P)^2-\frac{g^2P^2R}{r^2\Delta}=
 \frac{1}{2\Delta}\frac{\partial V}{\partial R}.
\end{equation}

The Einstein equation is also reduced to:
\begin{eqnarray}
 \frac{1}{2}\frac{\Delta'}{r}+\frac{1}{4}r^2e^{2\delta}(\Omega')^2&=&
 -\kappa^2\left[\frac{e^{2\delta}}{\Delta}g^2R^2(W+\Omega P)^2
 +\Delta(R')^2+\frac{(gPR)^2}{r^2}+V\right],\\
 \frac{\Delta}{r}\delta'&=&-2\kappa^2\left
 [\frac{e^{2\delta}}{\Delta}g^2R^2
 (W+\Omega P)^2+\Delta(R')^2\right],\\
 \frac{\Delta}{2r^2}(r^3e^{\delta}\Omega')'&=&
 2\kappa^2g^2R^2Pe^{\delta}\frac{W+\Omega P}{r}.
\end{eqnarray}

We set the metric which approaches asymptotically 
the BTZ solution when the scalar field falls 
into a vacuum of the negative value at radial infinity. 
In this setting, one can find that the cosmological 
constant $C$ is obtained to be 
$C=2(c^4+b^2+2bc^2)$ 
and as analyzed in~\cite{ref17}, the physical parameters ( the BTZ mass, 
the angular 
momentum and the magnetic flux) is the following:
\begin{eqnarray}
 M_{BTZ}&=&M(r_*)+\frac{r_*^4}{4}e^{2\delta(r_*)}(\Omega'(r_*))^2,\\
 J&=&r_*^3e^{\delta(r_*)}(\Omega'(r_*)), \\
 \Phi&=&2\pi P(r_*),
\end{eqnarray}
where $r_*$ is a sufficiently large value for radial coordinate $r$.

 These equations unfortunately cannot be solved analytically. Then we will 
find the numerical solution. In the next section, 
we discuss the relation among 
the physical parameters for various values of a cosmological constant.

%%%%%%%%%%%%%%%%%%%%%%%%%%%%%%%%%%%%%%%%%%%%%%%%%%%%%%%%%%%%%%%%%%%%%%
\section{Numerical result and Conclusion}
\label{sec:4}
%%%%%%%%%%%%%%%%%%%%%%%%%%%%%%%%%%%%%%%%%%%%%%%%%%%%%%%%%%%%%%%%%%%%%%
%%%%%%%%%%%%%%%%%%%%%%%%%%%%%%%%%%%%%%%%%%%%%%%%%%%%%%%%%%%%%%%%%%%%%%
In this paper, we have numerically constructed the non-topological 
solitions in three-dimensional $U(1)$-gauged $\mathcal{N}$$=2$ supergravity. 
The non-topological soliton does not preserve the supersymmetry, in general, 
because the metric is described by the BTZ solution at radial infinity. 

We have found that the region of the solutions can exist 
for the non-topological solitons in Fig.~1, shown with 
the BTZ mass $M_{BTZ}$, the angular momentum $J$ and the 
magnetic flux $\Phi$ for various values of the cosmological constant $C$. 
In Fig.~2, we find two branches in a sequence 
of the solutions which are obtained numerically 
by the different conditions ( values of the 
scalar field and the vector field at the origin). 
One branch is the solutions from the origin to 
the maximal value of the $M_{BTZ}$,
 the other is from the point $(0, -1)$ to the maximal value. 
As studied in~\cite{ref17}, one can also 
compare the results with black hole physics. 
In the case of the black hole, it is known that 
a black hole with a larger angular 
momentum than the critical one has a naked singularity 
and cannot be the usual black hole. 
Moreover there is the relation $\sqrt{C}J=M_{BTZ}$ 
which is satisfied by the extreme black hole. 
If the non-topological soliton becomes a black hole, 
the soliton with $\sqrt{C}J<M_{BTZ}$ 
can become the usual black hole. 
The non-topological soliton with $\sqrt{C}J>M_{BTZ}$ 
cannot be the usual black hole 
and the soliton with a large mass can certainly 
become an extreme black hole. 
Thus, the region of the solutions for the non-topological 
solitons is plausible as self-gravitating 
systems because all of the maximal values for 
the soliton's mass exist in the area $\sqrt{C}J<M_{BTZ}$. 
In Fig~3, if we fit the plotted line with a solid line, one can find 
that the $J$ is proportional to the square of the $\Phi$. This is the 
same results as in the Minkowski space-time case, 
as known in~\cite{ref13,ref14,ref15}. 
Therefore, the relation between the angular momentum and the magnetic flux 
is not dependent of the cosmological constant. 

There is a phase transition point between the symmetric and 
the asymmetric phase in 
the Abelian Chern-Simons Higgs model in Minkowski space-times. 
On the other hands, in the $U(1)$-gauged $\mathcal{N}$$=2$ supergravity, 
differed from the usual phase transition, there is not a 
non-topological soliton in the transition point because 
the metric approaches asymptotically anti de-Sitter space-time. 
It may be crucial in cosmological situation 
though this model has still three-dimensions.

In the present paper, we have treated the soliton with zero-winding number 
case, i.e. $n=0$ only. We will further investigate the binding energy, 
the stability of the soliton.

%%%%%%%%%%%%%%%%%%%%%%%%%%%%%%%%%%%%%%%%%%%%%%%%%%%%%%%%%%%%%%%%%%%%%%
\section*{Acknowledgement}
The authors would like to thank N.~Kan for useful advice.
%%%%%%%%%%%%%%%%%%%%%%%%%%%%%%%%%%%%%%%%%%%%%%%%%%%%%%%%%%%%%%%%%%%%%%
\begin{figure}
\epsfbox{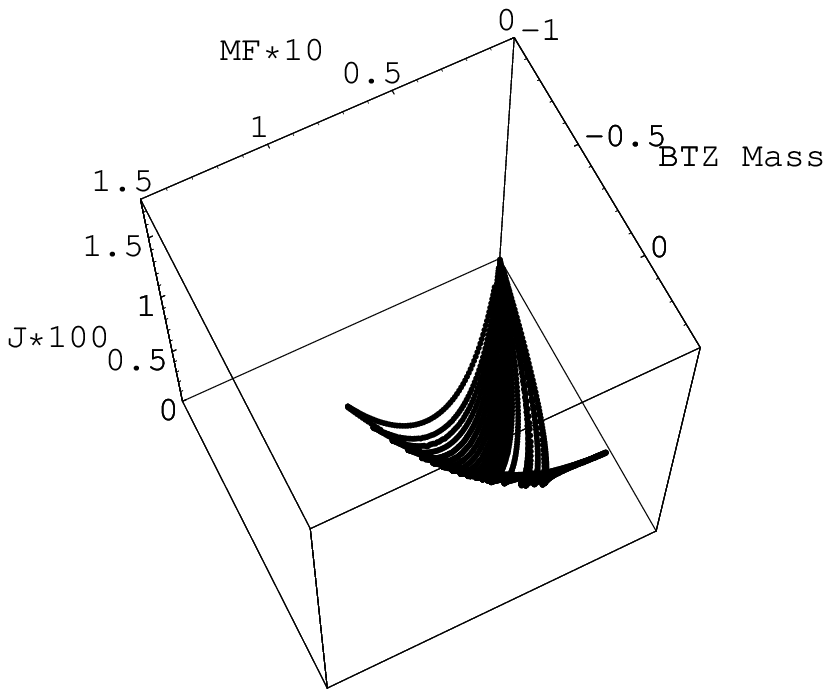}
\caption[Fig.~1]{The region of the non-topological 
soliton showing with the BTZ mass $M_{BTZ}$, 
the angular momentum $J$, and the magnetic flux 
$\Phi$ for various values of a cosmological constant $C$.
}
\epsfbox{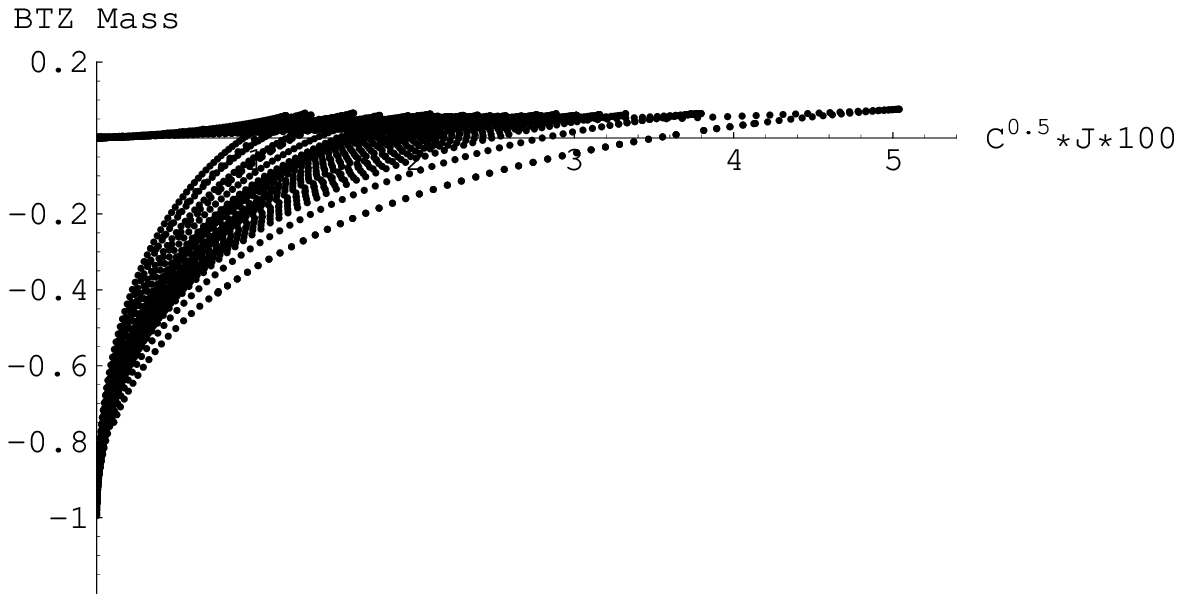}
\caption[Fig.~2]{The region of the non-topological soliton 
showing with $M_{BTZ}$ and $\sqrt{C}J$}

\epsfbox{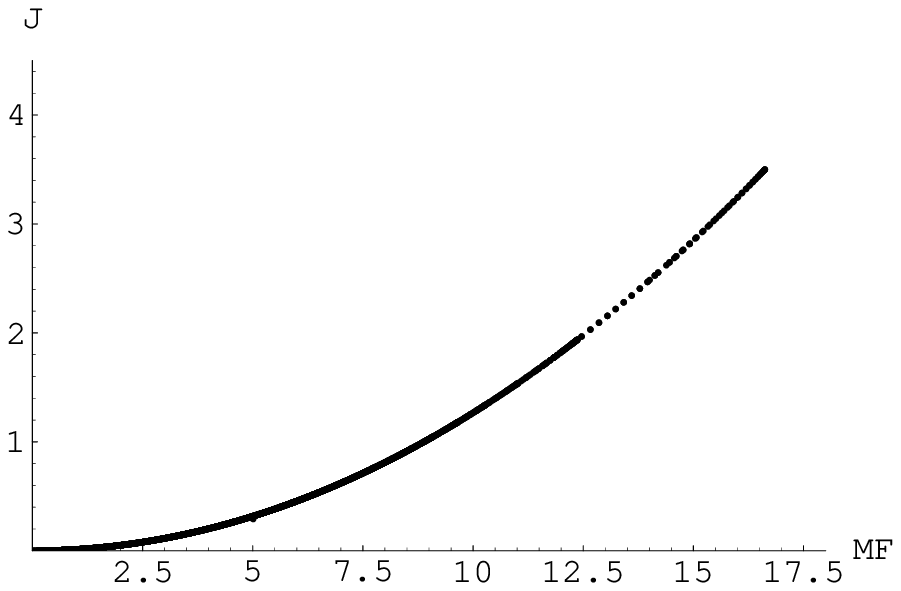}
\caption[Fig.~3]{The region of the non-topological 
soliton showing with $\Phi$ and $J$}
\end{figure}

%%%%%%%%%%%%%%%%%%%%%%%%%%%%%%%%%%%%%%%%%%%%%%%%%%%%%%%%%%%%%%%%%%%%%%
%%%%%%%%%%%%%%%%%%%%%%%%%%%%%%%%%%%%%%%%%%%%%%%%%%%%%%%%%%%%%%%%%%%%%%

%%%%%%%%%%%%%%%%%%%%%%%%%%%%%%%%%%%%%%%%%%%%%%%%%%%%%%%%%%%%%%%%%%%%%%
%%%References
%%%%%%%%%%%%%%%%%%%%%%%%%%%%%%%%%%%%%%%%%%%%%%%%%%%%%%%%%%%%%%%%%%%%%%

\newpage
%%%%%%%%%%%%%%%%%%%%%%%%%%%%%%%%%%%%%%%%%%%%%%%%%%%%%%%%%%%%%%%%%%%%%%
%%%FIGURES
%%%%%%%%%%%%%%%%%%%%%%%%%%%%%%%%%%%%%%%%%%%%%%%%%%%%%%%%%%%%%%%%%%%%%%

%%%%%%%%%%%%%%%%%%%%%%%%%%%%%%%%%%%%%%%%%%%%%%%%%%%%%%%%%%%%%%%%%%%%%%

\begin{references}

%1
\bibitem{ref1} 
S.~Deser,~R.~Jackiw,~and~G.~'t~Hooft,~Ann.~Phys.~(N.~Y.)~{\bf 152,}
~220~(1984).

%2
\bibitem{ref2} 
S.~Giddings,~J.~Abbott,~and~K.~Kuchar,~Gen.~Relativ.~Gravit.
~{\bf 16,}~751~(1984).

%3
\bibitem{ref3} 
E.~Witten,~Nucl.~Phys.~{\bf B311,}~46~(1988).

%4
\bibitem{ref4} 
I.~Bengtsson,~Phys.~Lett.~B~{\bf 220,}~51~(1989).

%5
\bibitem{ref5} 
S.~Carlip~and~J.~E.~Nelson,~Phys.~Lett.~B~{\bf 324,}~299~(1994).

%6
\bibitem{ref6} 
H.-J.~Matschull,~Class.~Quantum~Grav.~{\bf 16,}~2599~(1999).

%7
\bibitem{ref7} 
S.~Deser,~R.~Jackiw,~and~G.~'t~Hooft,
~Ann.~Phys.~(N.~Y.)~{\bf 153,}~405~(1984).

%8
\bibitem{ref8} 
S.~Deser~and~J.~H.~Key,~Phys.~Lett.~B~{\bf 120,}~97~(1983).

%9
\bibitem{ref9} 
A.~Ach\'{u}carro~and~P.~K.~Townsend,~Phys.~Lett.~B~{\bf 180,}~89~(1986).

%10
\bibitem{ref10} 
B.~de~Wit,~A.~K.~Tolls\'{e}n,~and~H.~Nicolai,~Nucl.~Phys.~{\bf B392,}
~3~(1993).

%11
\bibitem{ref11} 
H.~Nicolai~and~H.~Samtleben,~Phys.~Rev.~Lett.~{\bf 86,}~1686~(2001).

%12
\bibitem{ref12} 
H.~Nicolai~and~H.~Samtleben,~J.~High~Energy~Phys.~{\bf 04,}~022~(2001).

%16
\bibitem{ref16} 
M.~Abou-Zeid and~H.~Samtleben,~Phys.~Rev.~D~{\bf 65},~085016~(2002).%13

%13
\bibitem{ref13} 
J.~Hong,~Y.~Kim,~and~P.~Y.~Pac,~Phys.~Rev.~Lett.~{\bf 64,}~2230~(1990).

%14
\bibitem{ref14} 
R.~Jackiw~and~E.~J.~Weinberg,~Phys.~Rev.~Lett.~{\bf 64,}~2234~(1990).

%15
\bibitem{ref15} 
R.~Jackiw,~K.~Lee,~and~E.~J.~Weinberg,~Phys.~Rev.~D~{\bf 42,}~3488~(1990).

%19
\bibitem{ref19} 
E.~B.~Bogomol'nyi,~Sov.~J.~Nucl.~Phys.~{\bf 4},~449~(1976).

%17
\bibitem{ref17} 
K.~Sakamoto and~K.~Shiraishi,~Phys.~Rev.~D~{\bf 62},~124014~(2000).

%18
\bibitem{ref18} M.~Ba\~{n}ados, C.~Teitelboim and J.~Zanelli, 
Phys.\ Rev.\ Lett.\ {\bf 69}, 1849 (1992);
M.~Ba\~{n}ados, M.~Henneaux, C.~Teitelboim and J.~Zanelli, 
Phys.\ Rev.\ {\bf D48}, 1506 (1993).
For a review of BTZ black holes, S.~Carlip, Class.\ Quant.\ Grav.\ 
{\bf 12}, 2853 (1995) {\tt gr-qc/9506079}.

\end{references}
\end{document}